\documentstyle[12pt]{article}
\makeatletter
\newcommand{\singlespacing}{\let\CS=\@currsize\renewcommand{\baselinestretch}{1.0}\tiny\CS}
\newcommand{\doublespacing}{\let\CS=\@currsize\renewcommand{\baselinestretch}{1.5}\tiny\CS}

\begin{document}

\title{A PT-symmetric QES partner to the Khare-Mandal potential with real
eigenvalues}  

\author{B. Bagchi$^{a}$\thanks{{\bf E-mail : bagchi @ cucc.ernet.in}}, \ S.
Mallik$^{a}$, \ C. Quesne$^{b}$\thanks{{\bf E-mail : cquesne @
ulb.ac.be}} \ \& \ R. Roychoudhury$^{c}$\thanks{{\bf E-mail : raj @ 
www.isical.ac.in}}\\ \\ $^{a}$ Department of Applied Mathematics, University of
Calcutta,\\ 92 Acharya Prafulla Chandra Road, Kolkata 700009, India\\ \\ $^{b}$
Physique Nucl\'{e}aire Th\'{e}orique et Physique Math\'{e}matique,\\
Universit\'{e} Libre de Bruxelles, Campus de la Plaine CP229,\\ Boulevard du
Triomphe, B-1050 Brussels, Belgium\\ \\ $^{c}$ Physics \& Applied Mathematics
Unit,\\ Indian Statistical Institute,  203 B. T. Road,\\ Kolkata 700035,
India}  

\date{}

\maketitle

\centerline{{\bf ABSTRACT}}

\vspace*{0.2cm}

\thispagestyle{empty}

\setlength{\baselineskip}{18.5pt}

We consider a PT-symmetric partner to Khare-Mandal's recently proposed
non-Hermitian potential with complex eigenvalues. Our potential, which is
quasi-exactly solvable, is shown to possess only real eigenvalues.

\vspace*{0.3cm}

\noindent
{\bf PACS \ : \ 03.65.Bz, \ 03.65.Ge}

\vspace*{0.2cm}

\noindent
{\bf Keywords \ : \ quantum mechanics, PT-symmetry, quasi-solvable}

\vspace*{0.2cm}

\noindent
Corresponding author \ : \ B. Bagchi, Department of Applied Mathematics,
University of Calcutta, 92 Acharya Prafulla Chandra Road, Kolkata 700009

\noindent
Fax \ : \ (0091)(033)554 5741\\
E-mail \ : \ bbagchi @ cucc.ernet.in

\newpage

Exploration of non-Hermitian Hamiltonians, in particular the PT-symmetric
ones, is currently a topic of active research interest ( see for example,
[1]-[14]). As is well known, PT-symmetric Hamiltonians are conjectured [15] to
preserve the reality of their bound state eigenvalues except possibly for
situations when PT may be spontaneously broken. It should be noted that
PT-invariance in itself is not a sufficient condition for the Hamiltonian to
possess an entirely real spectrum [1, 2].

Recently, Khare and Mandal (KM) have enquired [9] into the invariance of a
non-Hermitian Hamiltonian under the combined operations of a complex shift ($x
\to a-x$, $a = i \frac{\pi}{2}$) and time reversal ($p \to -p$, $i \to -i$)
symmetries and have argued, by considering a specific model potential, that
the quasi-exactly solvable eigenvalues can emerge as complex conjugate pairs
if one of the potential parameters is an even integer or if it is an odd
integer and, in addition, the other potential parameter is large enough. They
also identify their complex shift with that of parity and as such look upon
their potential as the one enjoying PT-symmetry :

\begin{equation}
\displaystyle{V(x) ~=~ - \left( \zeta ~\cosh~ 2x - i M \right)^2}
\end{equation}
where the parameter $\zeta$ is real and $M$ is restricted to integer values only.

We wish to point out in this note that although the above potential is
invariant under the aforesaid transformations $x \to a-x$, and $i \to -i$ as
rightly claimed by KM, it is non-PT-invariant because with `{\it a}' imaginary
these transformations {\it do not} commute between themselves. We also observe
that the potential (1) does admit of a PT-symmetric partner namely 

\begin{equation}
\displaystyle{V(x) ~=~ - \left( \zeta ~\sinh~ 2x - i M \right)^2}
\end{equation}
(where, as in (1), $\zeta$ is real and $M$ an integer) which, as can be easily
checked, is invariant under the joint action of parity ($x \to -x$) and time
reversal ($i \to -i$). Further, it is quasi-exactly solvable as we demonstrate
below.

We begin by considering simultaneously the KM potential (1) and its modified
version (2). The corresponding Hamiltonians are ($\hbar = 2m = 1$) :

\begin{equation}
\displaystyle{H^{(+)} ~=~ - \frac{d^2}{dx^2} - \left( \zeta ~\cosh~ 2x - i M
\right)^2} 
\end{equation}

\begin{equation}
\displaystyle{H^{(-)} ~=~ - \frac{d^2}{dx^2} - \left( \zeta ~\sinh~ 2x - i M
\right)^2} 
\end{equation}
which can also be expressed together as

\begin{equation}
\displaystyle{H^{(\pm)} ~=~ - \frac{d^2}{dx^2} - \left[ \frac{\zeta}{2}
~(e^{2x} \pm e^{-2x}) - i M \right]^2 ~.} 
\end{equation}

Now, under a change of variable $x=\frac{1}{2}~\log~z$, $H^{(\pm)}$ become 

\begin{equation}
\displaystyle{H^{(\pm)} ~=~ - 4 z^2 \frac{d^2}{dz^2} - 4z \frac{d}{dz} -
\left[ \frac{\zeta}{2} ~ \left( z \pm \frac{1}{z} \right) - i M \right]^2 ~.} 
\end{equation}
As such if we set

\begin{equation}
\displaystyle{\mu^{(\pm)} (z) ~=~ z^{(1-M)/2} ~ e^{i \frac{\zeta}{4}~ (z \pm
\frac{1}{z})}}
\end{equation}
the Hamiltonians (6) can be mapped to their gauge-transformed forms 

\begin{equation}
\displaystyle{H_g^{(\pm)} ~=~ \left[ \mu^{(\pm)} (z) \right]^{-1} ~
H^{(\pm)} ~ \left[ \mu^{(\pm)} (z) \right] ~.}
\end{equation}

From the relations
\begin{equation}
\displaystyle{ \mu^{-1} ~\frac{d}{dz} \mu ~=~ \frac{d}{dz} +
\frac{\mu^\prime}{\mu}} 
\end{equation}

\begin{equation}
\displaystyle{ \mu^{-1} ~\frac{d^2}{dz^2} \mu ~=~ \frac{d^2}{dz^2} + 2
\frac{\mu^\prime}{\mu} ~\frac{d}{dz} + \frac{\mu^{\prime\prime}}{\mu}}
\end{equation}
where primes denote differentiations with respect to $z$, we easily obtain 

\begin{equation}
\begin{array}{lcl}
\displaystyle{ H_g^{(\pm)}} &=&\displaystyle{ - 4z^2 \frac{d^2}{dz^2} - 4z
\left( 2 \frac{\mu^\prime}{\mu} z + 1 \right) ~\frac{d}{dz} - 4z^2
\frac{\mu^{\prime\prime}}{\mu} - 4z \frac{\mu^\prime}{\mu}}\\
&&\displaystyle{- \left[ \frac{\zeta}{2} (z \pm \frac{1}{z}) - i M \right]^2
~.}\\ 
\end{array}
\end{equation}
Taking now into account the expressions
\begin{equation}
\displaystyle{ \frac{\mu^\prime}{\mu} ~=~ \frac{1 - M}{2z} + i \frac{\zeta}{4} ~
\left( 1 \mp \frac{1}{z^2} \right)}
\end{equation}

\begin{equation}
\displaystyle{ \frac{\mu^{\prime\prime}}{\mu} ~=~ \left( \frac{\mu^\prime}{\mu}
\right)^2 - \frac{1 - M}{2z^2} \pm i \frac{\zeta}{2}~ \frac{1}{z^3}}
\end{equation}
we arrive at the following representations of $H_g^{(\pm)}$ :

\begin{equation}
\displaystyle{ H_g^{(\pm)} =  - 4z^2 \frac{d^2}{dz^2} - \left[
2 i \zeta z^2 - 4 (M-2) z \mp 2 i \zeta \right]~\frac{d}{dz} + 2 i \zeta (M-1) z + 2
M - 1 \mp \zeta^2 ~.}
\end{equation}

We thus find that the Schr\"{o}dinger equation

\begin{equation}
\displaystyle{H^{(\pm)} ~ \psi^{(\pm)} (x) ~=~ E^{(\pm)} ~ \psi^{(\pm)} (x)}
\end{equation}
is equivalent to
\begin{equation}
\displaystyle{H_g^{(\pm)} ~ \phi^{(\pm)} (z) ~=~ E^{(\pm)} ~ \phi^{(\pm)} (z)}
\end{equation}
where $\displaystyle{\psi^{(\pm)} (x) = \mu^{(\pm)} (z) ~ \phi^{(\pm)} (z)}$.

It should be noted here that in terms of the $s\ell (2, R)$ generators [16]

\begin{equation}
\displaystyle{ J_+ ~=~ z^2 \frac{d}{dz} - 2 j z, ~ J_0 ~=~ z \frac{d}{dz} - j,
~  J_- ~=~ \frac{d}{dz}}
\end{equation}
the gauged Hamiltonians $H_g^{(\pm)}$ can be rewritten as
\begin{equation}
\displaystyle{ H_g^{(\pm)} ~=~ - 4 J_0^2 - 2 i \zeta J_+ \pm 2 i \zeta J_- + M^2
\mp \zeta^2}
\end{equation}
(provided $j = \frac{M-1}{2}$) in the true spirit of quasi-solvability[16].

We now turn to some specific cases of $\phi^{(\pm)} (z)$ by focussing on the
following choices :

\begin{enumerate}
\item[(i)] $\displaystyle{\phi^{(\pm)} (z) = c_0^{(\pm)}}$
\item[(ii)] $\displaystyle{\phi^{(\pm)} (z) = c_0^{(\pm)} + c_1^{(\pm)} z}$
$\qquad$ $\displaystyle{ \left( c_1^{(\pm)} \neq 0 \right)}$
\item[(iii)] $\displaystyle{\phi^{(\pm)} (z) = c_0^{(\pm)} + c_1^{(\pm)} z +
c_2^{(\pm)} z^2}$ $\qquad$ $\displaystyle{ \left( c_2^{(\pm)} \neq 0 \right)}$
\item[(iv)] $\displaystyle{\phi^{(\pm)} (z) = c_0^{(\pm)} + c_1^{(\pm)} z +
c_2^{(\pm)} z^2 + c_3^{(\pm)} z^3}$ $\qquad$ $\displaystyle{ \left(
c_3^{(\pm)} \neq 0 \right)}$ 
\end{enumerate}
where $c_i^{(\pm)}$ $(i = 0, 1, 2, 3)$ are constants. It is obvious that we
can generalize $\phi^{(\pm)} (z)$ to higher degrees of $z$ apart from the ones
chosen here.

First consider $\displaystyle{\phi^{(\pm)} (z) = c_0^{(\pm)}}$. For this case,
Eq.(16) becomes 
\begin{equation}
\displaystyle{ 2i\zeta (M-1) z + 2 M - 1 \mp \zeta^2 - E^{(\pm)} ~=~ 0}
\end{equation}
leading to 
\begin{equation}
\displaystyle{ 2 \zeta (M-1) ~=~ 0}
\end{equation}

\begin{equation}
\displaystyle{ 2 M -1 \mp \zeta^2 - E^{(\pm)} ~=~0 ~.}
\end{equation}
Hence $M=1$ $(j=0)$ and as a result
\begin{equation}
\displaystyle{E^{(\pm)} ~=~ 1 \mp \zeta^2 ~.}
\end{equation}
The accompanying wave functions read
\begin{equation}
\displaystyle{ \psi^{(+)} \propto ~~e^{\frac{i \zeta}{2}~ \cosh~2x}}
\end{equation}

\begin{equation}
\displaystyle{ \psi^{(-)} \propto ~~e^{\frac{i \zeta}{2}~ \sinh~2x} ~.}
\end{equation}
We therefore see that for $M=1$ the energy eigenvalues corresponding to (1) as
well as its modified PT-symmetric version (2) are real.

Next consider $\displaystyle{\phi^{(\pm)} (z) = c_0^{(\pm)} + c_1^{(\pm)} z}$.
Eq.(16) gives
\begin{equation}
\displaystyle{2i\zeta (M-2)~c_1^{(\pm)} ~=~0}
\end{equation}
\begin{equation}
\displaystyle{2i\zeta (M-1)~c_0^{(\pm)} + \left( 6M - 9 \mp \zeta^2 - E^{(\pm)}
\right) ~c_1^{(\pm)} ~=~0}
\end{equation}
\begin{equation}
\displaystyle{\pm 2i\zeta ~ c_1^{(\pm)} + \left( 2M - 1 \mp \zeta^2 - E^{(\pm)}
\right) ~c_0^{(\pm)} ~=~0 ~.}
\end{equation}
Eq.(25) implies $M=2$ while Eqs.(26) and (27) give
\begin{equation}
\displaystyle{2i\zeta ~c_0^{(\pm)} - \epsilon^{(\pm)} ~c_1^{(\pm)} ~=~0}
\end{equation}
\begin{equation}
\displaystyle{- \epsilon^{(\pm)} ~c_0^{(\pm)} \pm 2 i \zeta ~c_1^{(\pm)} ~=~0}
\end{equation}
where $\displaystyle{\epsilon^{(\pm)} = E^{(\pm)} - 3 \pm \zeta^2}$.

Solving (28) and (29) we get $\displaystyle{\epsilon_\pm^{(+)} = \pm 2 i \zeta}$
(complex) and $\displaystyle{\epsilon_\pm^{(-)} = \pm 2 \zeta}$ (real). We thus
find for $\displaystyle{M=2 ~ (j = \frac{1}{2})}$ the results
\begin{equation}
\displaystyle{ E_\pm^{(+)} ~=~ 3 \pm 2 i \zeta - \zeta^2}
\end{equation}

\begin{equation}
\displaystyle{ \psi_\pm^{(+)} \propto ~ e^{\frac{i \zeta}{2}~\cosh~2x}~(e^{-x} \pm
e^x)}
\end{equation}
for the KM potential and
\begin{equation}
\displaystyle{ E_\pm^{(-)} ~=~ 3 \pm 2 \zeta + \zeta^2}
\end{equation}

\begin{equation}
\displaystyle{ \psi_\pm^{(-)} \propto ~ e^{\frac{i \zeta}{2}~\sinh~2x}~(e^{-x} \pm
i e^x)}
\end{equation}
for the potential (2). Expectedly the eigenvalues for the KM case turn out to
be a complex conjugate pair, $M=2$ being an even integer. However, those for
the PT-symmetric potential (2) emerge real as borne out by (32).

Proceeding now to the case $\displaystyle{\phi^{(\pm)} (z) = c_0^{(\pm)} +
c_1^{(\pm)} z + c_2^{(\pm)} z^2}$, we find from Eq.(16)
\begin{equation}
\displaystyle{2 i \zeta (M-3)~c_2^{(\pm)} ~=~0}
\end{equation}

\begin{equation}
\displaystyle{2 i \zeta (M-2)~c_1^{(\pm)} + \left( 10 M - 25 \mp \zeta^2 -
E^{(\pm)} \right) ~c_2^{(\pm)} ~=~0}
\end{equation}

\begin{equation}
\displaystyle{2 i \zeta (M-1)~c_0^{(\pm)} + \left( 6 M - 9 \mp \zeta^2 -
E^{(\pm)} \right) ~c_1^{(\pm)} \pm 4 i \zeta c_2^{(\pm)}~=~0}
\end{equation}

\begin{equation}
\displaystyle{\pm 2 i \zeta ~ c_1^{(\pm)} + \left( 2M - 1 \mp \zeta^2 - E^{(\pm)}
\right) ~c_0^{(\pm)} ~=~0 ~.}
\end{equation}
Here we have $M=3$ and defining $\displaystyle{\epsilon^{(\pm)} = E^{(\pm)} -
9 \pm \zeta^2}$, we get
\begin{equation}
\displaystyle{ 2 i \zeta c_1^{(\pm)} - \left( \epsilon^{(\pm)} + 4 \right) ~
c_2^{(\pm)} ~=~0}
\end{equation}

\begin{equation}
\displaystyle{ 4 i \zeta c_0^{(\pm)} - \epsilon^{(\pm)} ~c_1^{(\pm)} \pm 4 i \zeta
c_2^{(\pm)} ~=~0}
\end{equation}

\begin{equation}
\displaystyle{- \left( \epsilon^{(\pm)} + 4 \right) ~ c_0^{(\pm)} \pm 2 i \zeta
c_1^{(\pm)} ~=~0 ~.}
\end{equation}

Solving (38)-(40) we obtain for $ M = 3$ $(j=1)$ the solutions 
\begin{equation}
\displaystyle{E_0^{(+)} ~=~ 5 - \zeta^2, ~ E_{\pm}^{(+)} ~=~ 7 - \zeta^2 \pm 2
\sqrt{~1-4 \zeta^2}} 
\end{equation}

\begin{equation}
\displaystyle{ \psi_0^{(+)} \propto ~e^{\frac{i \zeta}{2}~\cosh~2x}~ \sinh~2x}
\end{equation}

\begin{equation}
\displaystyle{ \psi_\pm^{(+)} \propto ~e^{\frac{i \zeta}{2}~\cosh~2x}~ \left[ 2
~\cosh~2x - \frac{i}{\zeta}~ \left( 1 \pm \sqrt{~1-4 \zeta^2} \right) \right]}
\end{equation}
for the KM potential and 
\begin{equation}
\displaystyle{E_0^{(-)} ~=~ 5 + \zeta^2, ~ E_{\pm}^{(-)} ~=~ 7 + \zeta^2 \pm 2
\sqrt{~1+4 \zeta^2}} 
\end{equation}

\begin{equation}
\displaystyle{ \psi_0^{(-)} \propto ~e^{\frac{i \zeta}{2}~\sinh~2x}~ \cosh~2x}
\end{equation}

\begin{equation}
\displaystyle{ \psi_\pm^{(-)} \propto ~e^{\frac{i \zeta}{2}~\sinh~2x}~ \left[ 2
~\sinh~2x - \frac{i}{\zeta}~ \left( 1 \pm \sqrt{~1 + 4 \zeta^2} \right) \right]}
\end{equation}
for the potential (2). Contrary to the Khare-Mandal potential (1) for which
two of the eigenvalues (namely $E^{(+)}_{\pm}$) become complex if $|\zeta|$ is
larger than the critical value $\zeta_c = \frac{1}{2}$, all three eigenvalues
of the PT-symmetric potential (2) remain real for all values of $\zeta$.

We now take up the case (iv) namely $\displaystyle{\phi^{(\pm)} (z) =
c_0^{(\pm)} + c_1^{(\pm)} z + c_2^{(\pm)} z^2 + c_3^{(\pm)} z^3}$ for which we
obtain from Eq.(16) the relations
\begin{equation}
\displaystyle{2 i \zeta (M-4)~c_3^{(\pm)} ~=~0}
\end{equation}

\begin{equation}
\displaystyle{2 i \zeta (M-3)~c_2^{(\pm)} + \left( 14 M - 49 \mp \zeta^2 -
E^{(\pm)} \right) ~c_3^{(\pm)} ~=~0}
\end{equation}

\begin{equation}
\displaystyle{2 i \zeta (M-2)~c_1^{(\pm)} + \left( 10 M - 25 \mp \zeta^2 -
E^{(\pm)} \right) ~c_2^{(\pm)} \pm 6 i \zeta c_3^{(\pm)}~=~0}
\end{equation}

\begin{equation}
\displaystyle{2 i \zeta (M-1)~c_0^{(\pm)} + \left( 6 M - 9 \mp \zeta^2 -
E^{(\pm)} \right) ~c_1^{(\pm)} \pm 4 i \zeta c_2^{(\pm)}~=~0}
\end{equation}

\begin{equation}
\displaystyle{\pm 2 i \zeta ~ c_1^{(\pm)} + \left( 2M - 1 \mp \zeta^2 -
E^{(\pm)} \right) ~c_0^{(\pm)} ~=~0 ~.}
\end{equation}
We are thus led to $M=4$ and defining $\displaystyle{\epsilon^{(\pm)} =
E^{(\pm)} - 15 \pm \zeta^2}$, we get
\begin{equation}
\displaystyle{ 2 i \zeta c_2^{(\pm)} - \left( \epsilon^{(\pm)} + 8 \right) ~
c_3^{(\pm)} ~=~0}
\end{equation}

\begin{equation}
\displaystyle{ 4 i \zeta c_1^{(\pm)} - \epsilon^{(\pm)} ~c_2^{(\pm)} \pm 6 i \zeta
c_3^{(\pm)} ~=~0}
\end{equation}

\begin{equation}
\displaystyle{ 6 i \zeta c_0^{(\pm)} - \epsilon^{(\pm)} ~c_1^{(\pm)} \pm 4 i \zeta
c_2^{(\pm)} ~=~0}
\end{equation}

\begin{equation}
\displaystyle{- \left( \epsilon^{(\pm)} + 8 \right) ~ c_0^{(\pm)} \pm 2 i \zeta
c_1^{(\pm)} ~=~0 ~.}
\end{equation}

An analysis of Eqs.(52)-(55) reveals that for consistency of the equations
$\epsilon^{(\pm)}$ have to fulfil the condition
\begin{equation}
\displaystyle{ \left( \epsilon^{(\pm)} + 8 \right) ~ \left[ \left(
\epsilon^{(\pm)} + 8 \right) ~\left( \epsilon^{(\pm) 2} \pm 16 \zeta^2 \right)
\pm 24 \zeta^2 \epsilon^{(\pm)} \right] + 144 \zeta^4 ~=~ 0 ~.}
\end{equation}
This fourth-degree equation can be factorized into two quadratic ones namely 
\begin{equation}
\displaystyle{\left( \epsilon^{(+)} + 8 \right)~\left( \epsilon^{(+)} \pm
4 i \zeta \right) + 12 \zeta^2 ~=~0}
\end{equation}

\begin{equation}
\displaystyle{\left( \epsilon^{(-)} + 8 \right)~\left( \epsilon^{(-)} \pm
4 \zeta \right) - 12 \zeta^2 ~=~0 ~.}
\end{equation}

Let us now consider the solutions of (57) and (58). It is obvious that in the
former case, all the four solutions are complex. This corresponds to the KM
scenario. However, in the latter case (which corresponds to the PT-symmetric
model (2)), we get quadratic equations in $\epsilon^{(-)}$ with real
coefficients : 
\begin{equation}
\displaystyle{ \epsilon^{(-) 2} + 4 (2 \pm \zeta) ~ \epsilon^{(-)} + 4 \zeta (\pm
8 - 3 \zeta) ~=~0 ~.}
\end{equation}
For the upper signs we obtain two real solutions for any $\zeta$ :

\begin{equation}
\displaystyle{ \epsilon_{+, \pm}^{(-)} ~=~- 2 (2 + \zeta) \pm 4 \sqrt{~1 - \zeta +
\zeta^2}} 
\end{equation}
and the same is true for the lower signs :
\begin{equation}
\displaystyle{ \epsilon_{-, \pm}^{(-)} ~=~ - 2 (2 - \zeta) \pm 4 \sqrt{~1 + \zeta +
\zeta^2} ~.} 
\end{equation}
Note that we can combine the four real solutions given by (60) and (61) in the
manner

\begin{equation}
\displaystyle{ \epsilon_{\sigma, \tau}^{(-)} ~=~- 2 (2 + \sigma \zeta) + 4 \tau
\sqrt{~1 - \sigma \zeta + \zeta^2}} 
\end{equation}
where $\sigma$, $\tau = +$, $-$.

Thus corresponding to the PT-symmetric potential (2) our findings for $M=4$
$(j = \frac{3}{2})$ are
\begin{equation}
\displaystyle{ E_{\sigma, \tau}^{(-)} ~=~ 11 - 2 \sigma \zeta + \zeta^2 + 4
\tau \sqrt{~1 - \sigma \zeta + \zeta^2}} 
\end{equation}

\begin{equation}
\displaystyle{ \psi_{\sigma, \tau}^{(-)} \propto ~ e^{\frac{i \zeta}{2}~
\sinh~2x}~ (e^{-x} - \sigma i e^x) ~ \left[ \sinh~2x - \frac{i}{\zeta}~ \left( 1 +
\tau \sqrt{~1 - \sigma \zeta + \zeta^2} \right) \right] ~.} 
\end{equation}

To summarize, we observe from the foregoing treatment of the cases $M=1, 2, 3,
4$ that unlike the case of KM potential in which QES eigenvalues occur in
complex conjugate pairs for $M$ an even integer but may be real for $M$ an odd
integer and $|\zeta|$ smaller than or equal to some critical value $\zeta_c$,
our PT-symmetric potential (2) exhibits real energy eigenvalues both 
for even and odd integer values of $M$ and any value of $\zeta$. It should be remarked that although we
restricted our discussion upto $M=4$ which corresponds to keeping a
fouth-degree wave function, it is clear that we can deal with, in an identical
way, higher degree contributions in $\phi^{(\pm)} (z)$. We conjecture that all
of them will lead to real eigenvalues.

Finally, we can write down recurrence relations for polynomials by
substituting\\ $\displaystyle{\phi^{(\pm)} (z) = \sum^\infty_{n=0}
\frac{R_n^{(\pm)} (E^{(\pm)})}{n !} t^n}$, $\displaystyle{t = \pm \frac{z}{2 i
\zeta}}$ in the Schr\"{o}dinger equation (16). It can be readily seen that the
coefficients $\displaystyle{R_n^{(\pm)} (E^{(\pm)})}$ satisfy the three-term
recursion relation 

\begin{equation}
\displaystyle{R_{n+1}^{(\pm)} ~\left( E^{(\pm)} \right)~=~ \left( E^{(\pm)} -
b_n^{(\pm)} \right)~R_n^{(\pm)} (E^{(\pm)}) - a_n^{(\pm)} R_{n-1}^{(\pm)}
(E^{(\pm)})} 
\end{equation}
where
\begin{equation}
\displaystyle{ a_n^{(\pm)} ~=~ \mp 4n ( M-n ) \zeta^2}
\end{equation}

\begin{equation}
\displaystyle{ b_n^{(\pm)} ~=~ 4n ( M-1-n ) + 2 M-1 \mp \zeta^2 ~.}
\end{equation}
If $M=k$, a positive integer, then $a_k^{(\pm)} = 0$ and for $n=k$, Eq.(65)
reduces to a two-term recursion relation. As a consequence, $R_{k+1}^{(\pm)}$,
and more generally $R_{k+n}^{(\pm)}$, is proportional to $R_k^{(\pm)}$ :

\begin{equation}
\displaystyle{R_{k+n}^{(\pm)} ~=~ R_k^{(\pm)} ~ \bar{R}_n^{(\pm)}}
\end{equation}
where $\bar{R}_n^{(\pm)}$ satisfies the three-term recursion relation

\begin{equation}
\displaystyle{\bar{R}_{n+1}^{(\pm)} ~\left( E^{(\pm)} \right)~=~ \left(
E^{(\pm)} - b_{M+n}^{(\pm)} \right)~\bar{R}_n^{(\pm)} (E^{(\pm)}) -
a_{M+n}^{(\pm)} ~\bar{R}_{n-1}^{(\pm)}~(E^{(\pm)}) ~.} 
\end{equation}
QES eigenvalues are obtained as solutions of the $k^{th}$-degree equation
$\displaystyle{R_k^{(\pm)} ~\left( E^{(\pm)} \right)~=~0}$.

SM thanks the Council of Scientific and Industrial Research, New Delhi for the
award of a fellowship. CQ is a Research Director of the National Fund for
Scientific Research (FNRS), Belgium.

\section*{References}

\begin{enumerate}
\item[[1]] C. M. Bender, S. Boettcher : Phys. Rev. Lett {\bf 80} (1998)
5243. 
\item[[2]] C. M. Bender, S. Boettcher : J. Phys {\bf A 31} (1998) L 273.
\item[[3]] E. Delabaere, F. Pham : Phys. Lett {\bf A 250} (1998) 25, 29.
\item[[4]] M. Znojil :  Phys. Lett {\bf A 259} (1999) 220.
\item[[5]] B. Bagchi, R. Roychoudhury : J. Phys {\bf A 33} (2000) L 1.
\item[[6]] M. Znojil : J. Phys {\bf A 33} (2000) L 61.
\item[[7]] M. Znojil : J. Phys {\bf A 33} (2000) 4561.
\item[[8]] B. Bagchi, F. Cannata, C. Quesne : Phys. Lett {\bf A 269} (2000)
79. 
\item[[9]] A. Khare, B. P. Mandal : Phys. Lett {\bf A 272} (2000) 53. 
\item[[10]] B. Bagchi, C. Quesne : Phys. Lett  {\bf A 273} (2000) 285. 
\item[[11]] F. Cannata, M. Ioffe, R. Roychoudhury, P. Roy : Phys. Lett  {\bf
A 281} (2001) 305. 
\item[[12]] G. L\'{e}vai, F. Cannata, A. Ventura : J. Phys {\bf A 34} (2001)
839. 
\item[[13]] Z. Ahmed : Phys. Lett {\bf A 282} (2001) 343. 
\item[[14]] P. Dorey, C. Dunning, R. Tateo : J. Phys. {\bf A  34} (2001) L391.
\item[[15]] D. Bessis : Unpublished (1992).
\item[[16]] A. Ushveridze : Quasi-Exactly Solvable Models in Quantum Mechanics,
Institute of Physics Publishing, Bristol (1994).
\end{enumerate} 

\end{document}